\documentclass[sigconf]{acmart}
\AtBeginDocument{%
  }

\copyrightyear{2026}
\acmYear{2026}
\setcopyright{cc}
\setcctype{by}
\acmConference[ICSE-FoSE]{2026 IEEE/ACM 48th International Conference on Software Engineering: Future of Software Engineering }{April 12--18, 2026}{Rio de Janeiro, Brazil}
\acmBooktitle{2026 IEEE/ACM 48th International Conference on Software Engineering: Future of Software Engineering (ICSE-FoSE), April 12--18, 2026, Rio de Janeiro, Brazil}
\acmPrice{}
\acmDOI{10.1145/3793657.3793873}
\acmISBN{979-8-4007-2479-4/2026/04}

\usepackage{graphicx}
\usepackage{float}
\begin{document}

\title{How Software Engineering Research Overlooks Local Industry: A Smaller Economy Perspective}

\author{Klara Borowa}
\email{klara.borowa@pw.edu.pl}
\affiliation{%
  \institution{Warsaw University of Technology, Institute of Control and Computation Engineering}
  \city{Warsaw}
  \country{Poland}
}

\author{Andrzej Zalewski}
\affiliation{%
  \institution{Warsaw University of Technology, Institute of Control and Computation Engineering}
  \city{Warsaw}
  \country{Poland}
}

\author{Lech Madeyski}
\email{lech.madeyski@pwr.edu.pl}
\affiliation{%
  \institution{Wroclaw University of Science and Technology, Faculty of Information and Communication Technology}
  \city{Wroclaw}
  \country{Poland}
}

\renewcommand{\shortauthors}{Borowa et al.}

\begin{abstract}
 The software engineering researchers from countries with smaller economies, particularly non-English speaking ones, represent valuable minorities within the software engineering community. As researchers from Poland, we represent such a country.   We analyzed the ICSE FOSE (Future of Software Engineering) community survey through reflexive thematic analysis to show our viewpoint on key software community issues. We believe that the main problem is the growing research-industry gap, which particularly impacts smaller communities and small local companies. Based on this analysis and our experiences, we present a set of recommendations for improvements that would enhance software engineering research and industrial collaborations in smaller economies.
\end{abstract}




\maketitle

\section{Introduction and Background}


This paper has been inspired by the ICSE FOSE community survey~\cite{storey_2025_18217799}~\footnote{https://conf.researchr.org/track/icse-2026/icse-2026-future-of-software-engineering}. We, Polish software engineering (SE) researchers of varying levels, strived to use these surveys' results to understand the current worldwide SE community better and identify whether our view matches with it. 

Our standpoint is one of researchers from a smaller economy, where English is not the dominant language. Each of us represents a different academic career stage in Poland, from first to third author respectively: an early-career researcher who recently completed a PhD in SE with early exposure to top-tier SE venues (ICSA, ECSA, JSS, IEEE SW), a mid-career researcher with moderate exposure to top-tier SE venues (ICSA, ECSA, JSS, CSUR, IEEE SW), and a senior researcher with more extensive exposure to top-tier SE venues (ICSE, ASE, FSE, TSE, IST, EMSE, JSS, IEEE SW). This positionality provides us with firsthand experience of the barriers that researchers from smaller economies face, while also shaping our sensitivity to these issues.
Polish researchers rarely participate in ICSE, with only a few instances in the last years according to Scopus. This standpoint also matches many other countries, which could possibly provide more diverse data and insights if included.

As such, this paper's \textbf{goal} is to propose a set of changes that would benefit SE researchers from smaller economies. 
In this work, we analyzed the community survey through reflexive thematic analysis~\cite{braun2022thematic}(Section~\ref{sec:method}), to understand the current SE community's standpoint on its strong points, weak points, and possible solutions. These were gathered into themes showcasing community points of focus (Section~\ref{sec:themes}). We used them to create an overarching theme (Section~\ref{sec:overarching_theme}) representing one key problem for our SE community: perpetuating the research-industry gap. We argue that this gap has a stronger hold on countries with smaller economies, such as Poland, where smaller domestic software companies operate.

Research on the SE research-industry gap has been limited so far. 
Rarely, researchers publish papers showcasing gaps and potential areas of research based on industry needs.
Begel and Zimmermann~\cite{begel_analyze_2014} surveyed data scientists to obtain topics worth researching. 
To find gaps between research and practice, Kochhar et al.~\cite{kochhar_practitioners_2016} surveyed practitioners to find out how fault localization research impacts them. Stradowski and Madeyski~\cite{stradowski_bridging_2023} explicitly researched the research-industry gap in machine learning software defect prediction, and proposed a set of key considerations for researchers.

This key issue has garnered some attention lately. Winters has written a letter to the JSS Editor where he states clearly ``If we want to make research venues like JSS generally applicable to industry practitioners, the density of papers that help with common real-world tasks needs to increase.''~\cite{winters2024}.
Additionally, at ICSA (International Conference on Software Architecture) 2025, a joint effort was undertaken to create an industry-researcher collaboration manifesto, the Odense Manifesto~\cite{jorgensen_dear_2026}, which outlines key points of focus for joint research with industry.

We took into account the ICSE community survey, our experience as researchers from a smaller economy, and existing literature on the research-industry gap and proposed a set of solutions, which are presented in Section~\ref{sec:vision}.








\section{Method}
\label{sec:method}




We used the publicly available ICSE 2026 community survey as our base dataset. It includes responses from 280 software engineering community members, mainly tenured faculty from Europe, most with over 11 years of experience (see Table 1 in the Appendix~\cite{borowa_2026_18340061} for demographic details).


For our analysis, we used the following open-ended survey questions: (7) What aspect or aspects of the software engineering research community work well, and why? (8) What aspect or aspects of the software engineering research community do not work well, and why? (9) If you could make one change, what would you change, and what outcome from that change would you like to see in the software engineering research community?

We chose these questions for analysis because they represented three major focal points: ``the good'' strong points according to the community, ``the bad'' weak points according to the community, and ``the solutions'' that the SE community members proposed.

To understand the global SE community view, we employed reflexive thematic analysis, using the guidelines by Braun and Clarke~\cite{braun2022thematic}. We employed this method due to its theoretical flexibility and the reflexivity it promotes. This way, we could engage with the community survey data not only to find out what participants believe, but also to use our standpoint as Polish SE researchers to focus on themes crucial to our community, which is a minority within the global SE community.
Our analysis included the following steps:

\noindent\textbf{(1) Familiarization with the data:} We read the answers from the community survey to get an initial understanding of the data.

\noindent\textbf{(2) Initial coding}: The first author manually coded data segments from the questionnaire answers, creating a set of initial codes. E.g. ``The obvious solution is to go toward a more 'journal-oriented' publication model as pioneered by FSE (or other domains before ACM SIGSOFT), where conference attendance is no longer mandatory. '' was marked with the code ``Todo\#journal first focus''. ``Todo'' was a marker for proposed change, and ``journal first focus'' was a code used for the solution idea to make conferences focus on pre-published journal papers.
    
\noindent\textbf{(3) Reflexive iterative coding refinement:} We discussed the codes and their relationships during subsequent meetings, and the coding was iteratively refined into the ultimate code list. E.g. codes ``Todo\#focus on journals'', ``Todo\#only journal first'' and ``Todo\#more journal first'' were all merged into ``Todo\#more journal first'', since all suggested that putting more focus on the journal first track is possibly the solution to various conference problems (like big reviewer workload or low quality peer review).

\noindent\textbf{(4) Sub-theme generation:} We gathered the codes into sub-themes. Based on the codes, we created sub-theme wordclouds (see Figure~\ref{fig:big_figure}).

\noindent\textbf{(5) Theme generation:} We gathered the sub-themes into themes that represented the main areas discussed by the participants.

\noindent\textbf{(6) Overarching theme creation:} We discussed the themes and sub-themes during a meeting to establish which ones were of interest to the Polish SE community. Based on that, we chose a pair of sub-themes that became the overarching theme from our perspective, i.e. ``Perpetuating the research-industry gap''. 

During coding and sub-theme generation, we split codes/sub-themes into categories ``Good'', ``Bad'' and ``Todo'', representing: the community strong points, its weaknesses, and improvement ideas. 
We also split improvement ideas into actionable and non-actionable during coding refinement, and only included actionable solutions in the results. Non-actionable ideas represented the need for change, without proposing tools on how to achieve it, e.g. ``MAKE reviewers heart open to adopt new technologies used in research.''

\section{Results}
\begin{figure*}[h]
  \centering
  \includegraphics[width=\linewidth]{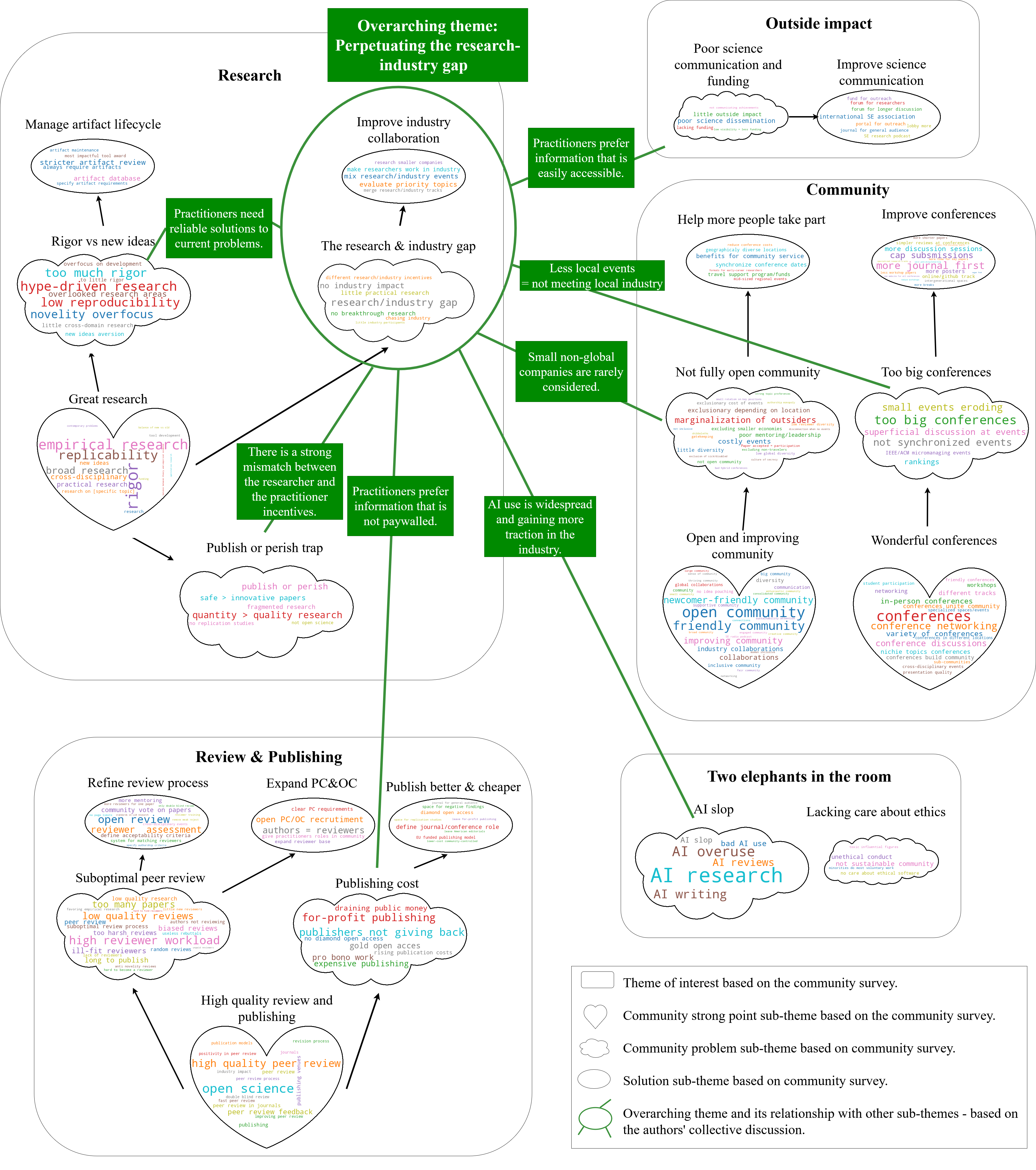}
  \caption{Theme map }
  \label{fig:big_figure}
    \Description[Theme map]{This image showcases a set of wordclouds for sub-themes. The sub-themes gather together into themes. The overarching theme "perpetuating the research-industry gap" is showed as well - and its connection with various subthemes.}
\end{figure*}

In this section, we present the themes from the community survey, and then showcase how the overarching theme ``Perpetuating the research-industry gap'' is a key area from our standpoint, with focus on the Polish SE community's perspective.
Figure~\ref{fig:big_figure} showcases all sub-themes, themes, and the overarching theme. 

The themes are presented in Section~\ref{sec:themes} in a shortened manner. Detailed theme descriptions with participant quotations are available in the Appendix~\cite{borowa_2026_18340061}. 

\subsection{Themes}
\label{sec:themes}

\noindent\textbf{Theme: Research}


The community seems to take great pride in its research. 
Particularly, participants praise quality empirical work, reproducibility, and methodological rigor.

However, community members have contradictory stances regarding rigor and new ideas. 
Many believe that currently rigor is excessive and makes it impossible to publish about new promising ideas. Yet, others hold the opposite view and would prefer even more rigor than currently. The codes ``hype-driven focus'', ``low reproducibility'' and ``novelty overfocus'' represent the community part that would prefer to focus on methodological rigor, while ``too much rigor'' represents the ones that would rather focus on exploring new promising ideas.
One solution that participants proposed can be summarized as having a clear artifact management lifecycle. This included stricter artifact evaluation, long-term artifact maintenance, and dedicated databases for these artifacts.

Another problem that strongly impacts the community appears to be the ``publish or perish'' effect.
Researchers are pressured to publish often in order to be promoted, keep their positions, and obtain funding. This pressure causes them to focus on the quantity of published papers rather than conducting valuable research. No actionable solutions were proposed in the community survey for this problem. Solutions mainly focused on ``changing minds'' in unspecified ways.

Finally, participants report the troubling research/industry gap, which was described by one participant as ``The so-called 'SE research community' is far from real issues and problems found by practitioners.'' 
Researchers often chase the industry to learn from it, and cannot keep up with current industry practices and challenges.
Additionally, collaborations, when they occur, are undervalued -- papers from industrial tracks in most conferences are considered less prestigious.
To address this gap, participants proposed several ideas. 
One dominating idea was joint events with more industry participants. Another was to regularly evaluate priority topics for practitioners and focus our research on them. Some even proposed requiring researchers to work in industry on a regular basis.





\noindent\textbf{Theme: Review \& Publishing}


Overall, the community seems to take pride in its high-quality review and publishing practices. Reviews are usually considered to be of good quality, and SE publishing venues are well regarded.


However, many participants felt that review quality is not optimal.
The main reason for that seems to be due to high reviewer workloads -- one participant called it ``Endless reviewing''.
Two types of solutions were proposed to address this issue. One of them was to expand Program Committees by having authors participate in reviewing, and doing open calls for committee members.
Secondly, refining the review process by using the open review platform and creating a mechanism for assessing review quality.



Another problem mentioned was the publishing cost. Participants felt that publishers do not ``give back'' enough to the community to justify these costs. 
One solution proposed would be redefining journal/conference roles in publishing, and publishing more mature work in journals.
Another notable example would be implementing the diamond open-access model. 




\noindent\textbf{Theme: Community}

Participants overwhelmingly believed that the SE community is open and friendly. 
Many also noted that the community has been self-improving in many aspects, e.g., by creating methodological guidelines.
Additionally, conferences were noted by numerous participants as the best aspect of the community, where collaboration and idea exchange can flourish.




However, many participants felt that some groups, particularly researchers from smaller countries, face more hardships in attempting to participate in community events. This stems from the lower income-to-participation cost ratio and visa restrictions.
Proposed solutions for this problem included discounts for reviewing and volunteering, 
and creating a travel support program. 


Another problem was that conferences became too big, which
results in fast and short presentations with little discussion/networking due to the hurry.
Two main solution ideas to this were capping submission numbers and limiting the number of tracks with new papers (with more focus on journal first tracks).




\noindent\textbf{Theme: Outside Impact}

Another issue pointed out in the survey was the vicious cycle resulting from poor science communication. 
Practitioners rarely hear about SE findings, and having such small visibility negatively impacts SE research funding.
To solve this, participants proposed funding outreach initiatives and active lobbying at the EU level.



Two additional issues emerged as purely negative with no proposed solutions: `AI slop'' (AI use in research, writing, and reviews) and lack of attention to ethical issues (see Appendix~\cite{borowa_2026_18340061}).

\subsection{Overarching theme: perpetuating the research-industry gap}
\label{sec:overarching_theme}

During the authors' discussion, we found that the research-industry gap is a systematic problem that impacts us in unexpected ways:

\noindent (1) Practitioners need reliable solutions to current problems. 
Extreme rigour is too slow, while unproven new ideas'' are untrustworthy, which naturally creates a two-speed world, making research ill-suited for industry. 
This aligns with S5: Relevance over originality'' and ``S6: Relevance over rigour'' in~\cite{jorgensen_dear_2026}, highlighting the need for just-in-time reliable solutions.


\noindent (2) Due to researchers' ``publish or perish'' focus, priorities are a mismatch with practitioners, who put little value on publishing papers. What practitioners care about is the timely delivery of valuable software. The Odense Manifesto mentions this issue as well, suggesting that it must be stopped (``S4: Practical applications of knowledge over smallest publishable unit''). In Poland, this pressure is tremendous. Our institutions are evaluated by our government regularly, and papers are assigned ``points''. Gathering these points is crucial for obtaining funding and promotion, creating a formal country-wide ``publish or perish'' system.

\noindent (3) Practitioners have many free information sources and rarely consult research papers~\cite{josyula2018software}. Even practitioner-oriented magazines like IEEE Software require subscriptions---there is no open platform for SE research dissemination. In Poland, only practitioners in large international companies have access to repositories like IEEE Xplore, and typically to just one.

\noindent (4) Due to the pressure to publish in large venues such as ICSE, FSE and ASE, small local conferences are disappearing. These were a powerful community-building tool for Polish researchers and practitioners that disappeared.
One such example is KKIO, the main Polish Software Engineering conference~\footnote{https://kkio.pti.org.pl/}. Between 1999 and 2023, it was organised in Poland and tried to unite our research and (to some extent) industry communities. However, due to the pressure to publish in ``better venues'' to deal with the ``publish or perish'' system, KKIO decided to join (and become a sub-event of) an established SE conference--- Euromicro DSD/SEAA. This may have enhanced its prestige, but in practice meant that Polish researchers may not consider it as a main gathering event anymore.

\noindent (5) One type of company has been systematically omitted - small local companies. While having young or less wealthy researchers attending conferences is a challenge, we noticed that no one was suggesting inviting small companies. These companies play a significant role in the Polish IT industry, yet they have been overlooked in favor of large corporations. An example of an overlooked area is Game Software Development, a topic around which 12 papers are published yearly~\cite{chueca2024consolidation}. Poland has a large number of small indie game studios that could be a rich source of information. 

\noindent (6) AI use is rapidly adopted in industry, yet the SE community often views AI research as problematic. We believe the influx of AI-related papers stems from real industry needs, not a temporary trend. We experience this duality ourselves, while industry makes AI proficiency a requirement for employment. In Poland, we have observed a steady decline in junior positions as companies use AI for simple tasks; recruitment events at our universities have notably declined in recent years. \textbf{We argue that the community's tendency to dismiss AI research as 'slop' is a manifestation of the research-industry gap: rejecting steps towards solving a top industrial need because the steps are not ``interesting for me''.} 

\section{Vision Discussion}
\label{sec:vision}
         

To address the overarching theme and aid smaller-economy researchers, we propose solutions to improve research and bridge the research-industry gap:

\noindent  \textbf{(1) Local conference revival program. } To participate in the software engineering community, researchers from smaller economies need conferences that are geographically close to them and affordable. However, simply organizing them with local researchers would result in ``not prestigious enough'' conferences where no one is willing to publish. 
We prompt the SE community to create an association to gather SE conference organizers. Such a group could synchronize top conference locations across continents, targeting smaller economy countries, and create tracks where organizers of local SE events could consult senior community members and share challenges and solutions that worked for smaller communities.

\noindent  \textbf{(2) Small company inclusion initiative. } Since small local companies are being overlooked, we should invite them and create collaboration incentives. As such, we envision a worldwide program targeting small companies, where researchers can access information about these companies' problems and work together to solve them. 
This could be done by researchers contacting local associations of small and mid-sized companies. 
One such organization, the third author of this paper had an opportunity to get in touch with, is ITCorner \footnote{https://itcorner.org.pl/en/about-us/who-we-are/}.

\noindent  \textbf{(3) Practitioner-accessible publishing. } We believe that it is possible for the members of the SE community to create platforms with diamond open access -- i.e., with no payment required from authors and readers. The third author of this paper was the driving force behind establishing the e-Informatica Software Engineering Journal (EISEJ)\footnote{https://www.e-informatyka.pl/} with crucial support from many distinguished members of its Editorial Board. EISEJ is based on volunteer work from community members and supported by the Department of Artificial Intelligence, Wroclaw University of Science and Technology. It is indexed by all major databases, including Web of Science (IF 2024 = 1.2) and Scopus (CiteScore 2024 = 3.5, CiteScoreTracker 2025 = 4.0), and is attracting submissions focused on SE and data science/AI/ML in SE.
With the high engagement and skill set of the current SE community members, this model may be usable on a larger scale.

\noindent \textbf{(4) AI research integration: Dedicated tracks with appropriate PC expertise. } In our opinion, research on AI in SE (AI4SE) and SE for AI (SE4AI) is absolutely necessary and, due to its high relevance to industry, cannot be overlooked. Since the current peer-review model at SE conferences often requires AI papers to be reviewed by reviewers who dislike or lack knowledge about this topic, dedicated tracks could be established. At such tracks, only reviewers interested in the topic would form the core of the PC. AI researchers from outside of our community should be invited to the PC of these tracks as well, to help address the current skill gap that the SE community is experiencing. 
Finally, we would like to note: \textbf{We strongly believe that excluding AI research from SE would be a grave mistake, one that would make the research-industry gap much more severe than it is currently}. 

\noindent \textbf{(5) Integrate AI into teaching SE. }
Since the widespread use of AI seems inevitable, both in professional and academic contexts, we believe that universities should reconsider their teaching methods in teaching SE.
Simply forbidding AI use is very unlikely to positively impact either the students or the teachers. Students would simply wonder why they cannot use typical tools widely available in the industry, and teachers would be forced to waste effort on detecting AI use.
Instead, students could perform more complex tasks with the use of AI. For example, instead of manually implementing a basic algorithm, they could generate and assess the results of many implementation variants that solve the same problem. 

An example of such an educational course is ``Research and Development Project in Software Engineering", where students in small teams (supervised by the third author of this paper) conduct an R\&D project that heavily focuses on AI in SE, starting from identifying and trying to reproduce state-of-the-art (SOTA) solutions and then proposing better approaches to the analyzed problems.   

\section{Conclusion}
We analyzed the ICSE FOSE community survey~\cite{storey_2025_18217799} and complemented it with our perspective as researchers from a smaller economy. We found that the most pressing issue is the growing research-industry gap, which impacts us strongly, in terms of the disconnect from the realities of our small local companies. 
In Section~\ref{sec:vision}, we present our vision for changes that would improve the functioning of SE researchers from smaller economies and their local collaboration with software practitioners working in small companies.

\noindent\textbf{Data availability} We include a replication package online~\cite{borowa_2026_18340061}.

\bibliographystyle{ACM-Reference-Format}
\bibliography{refs}

\end{document}